\documentclass{amsart}
\usepackage{amsmath}
\usepackage{graphicx}
\usepackage{amsfonts}
\usepackage{amssymb}
\newtheorem{theorem}{Theorem}
\theoremstyle{plain}

\newtheorem{definition}{Definition}
\newtheorem{example}{Example}

\newtheorem{proposition}{Proposition}
\newtheorem{remark}{Remark}

\numberwithin{equation}{section}
\begin{document}
\title[Bachelier versus Black-Scholes]{How close are the option pricing formulas of Bachelier and Black-Merton-Scholes?}
\author{Walter Schachermayer and Josef Teichmann}
\address{Institute of Mathematical Methods in Economics, Vienna University of
Technology, Wiedner Hauptstrasse 8-10, A-1040 Vienna, Austria}
\email{wschach@fam.tuwien.ac.at, jteichma@fam.tuwien.ac.at}

\begin{abstract}
We compare the option pricing formulas of Louis Bachelier and
Black-Merton-Scholes and observe -- theoretically as well as for Bachelier's
original data -- that the prices coincide very well. We illustrate Louis
Bachelier's efforts to obtain applicable formulas for option pricing in
pre-computer time. Furthermore we explain -- by simple methods from chaos
expansion -- why Bachelier's model yields good short-time approximations of
prices and volatilities.

\end{abstract}
\maketitle

\section{Introduction}

It is the pride of Mathematical Finance that L. Bachelier was the first to
analyze Brownian motion mathematically, and that he did so in order to develop
a theory of option pricing (see \cite{Bac:00}). In the present note we shall
review some of the results from his thesis as well as from his later textbook
on probability theory (see \cite{Bac:12}), and we shall work on the remarkable
closeness of prices in the Bachelier and Black-Merton-Scholes model.

The \textquotedblleft fundamental principle\textquotedblright\ underlying
Bachelier's approach to option pricing is crystallized in his famous dictum
(see \cite{Bac:00}, p.34)

\[
\text{\textquotedblleft L'\'{e}sperance mathematique du sp\'{e}culateur est
nul\textquotedblright,}
\]
i.e. \textquotedblleft the mathematical expectation of a speculator is
zero\textquotedblright. His argument in favor of this principle is based on
equilibrium considerations (see \cite{Bac:00} and \cite{Sch:00}), similar to
what in today's terminology is called the \textquotedblleft efficient market
hypothesis\textquotedblright\ (see \cite{Sam:65}), i.e. the use of martingales
to describe stochastic time evolutions of price movements in ideal markets. L.
Bachelier writes on this topic (see the original french version in
\cite{Bac:00}, p. 31).
\begin{align*}
&  \text{\textquotedblleft It seems that the market, the aggregate of
speculators, can believe }\\
&  \text{in neither a market rise nor a market fall, since, for each }\\
&  \text{quoted price, there are as many buyers as sellers.\textquotedblright}
\end{align*}

The reader familiar with today's approach to option pricing might wonder where
the concepts of \textquotedblleft risk free interest rate\textquotedblright
\ and "risk neutral measure" have disappeared to, which seem crucial in the
modern approach of pricing by no arbitrage arguments (recall that the
\emph{discounted} price process should be a martingale under the risk neutral
measure). As regards the first issue L. Bachelier applied his
\textquotedblleft fundamental principle\textquotedblright\ in terms of
\textquotedblleft true\textquotedblright\ prices (this is terminology from
1900 which corresponds to the concept of forward prices in modern
terminology), since all the payments involved (including the premium of the
option) were done only at maturity of the contracts. See \cite{Bac:00} for an
explicit description of the trading rules at the bourse de Paris in 1900. It
is well-known that the passage to forward prices makes the riskless interest
rate disappear: in the context of the Black-Merton-Scholes formula, this is
what amounts to the so-called Black's formula (see \cite{Bla:76}). As regards
the second issue L. Bachelier apparently believed in the martingale measure as
the historical measure, i.e. for him the risk neutral measure conincides with
the historical measure. For a discussion of this issue compare \cite{Sam:65}.

\bigskip

Summing up: Bachelier's \textquotedblleft fundamental
principle\textquotedblright\ yields $\emph{exactly}$ the same recipe for
option pricing as we use today (for more details we refer to the first section
of the St. Flour summer school lecture \cite{Sch:00}): using forward prices
(\textquotedblleft true prices\textquotedblright\ in the terminology of 1900)
one obtains the prices of options (or of more general derivatives of European
style) by taking expectations. The expectation pertains to a probability
measure under which the price process of the underlying security (given as
forward prices) satisfies the fundamental principle, i.e. is a martingale in
modern terminology.

It is important to emphasize that, although the \emph{recipes} for obtaining
option prices are the same for Bachelier's as for the modern approach, the
\emph{arguments} in favour of them are very different: an equilibrium argument
in Bachelier's case as opposed to the no arbitrage arguments in the
Black-Merton-Scholes approach. With all admiration for Bachelier's work, the
development of a theory of hedging and replication by dynamic strategies,
which is the crucial ingredient of the Black-Merton-Scholes-approach, was far
out of his reach (compare \cite{Sch:00} and section 2.1 below).

\bigskip

In order to obtain option prices one has to specify the underlying model. We
fix a time horizon $T>0$. As is well-known, Bachelier proposed to use
(properly scaled) Brownian motion as a model for forward stock prices. In
modern terminology this amounts to
\begin{equation}
S_{t}^{B}:=S_{0}+\sigma^{B}W_{t},\label{bacheliermodel}
\end{equation}
for $0\leq t\leq T$, where $(W_{t})_{0\leq t\leq T}$ denotes standard Brownian
motion and the superscript $B$ stands for Bachelier. The parameter $\sigma
^{B}>0$ denotes the volatility in the Bachelier model. Notice that in contrast
to today's standard Bachelier measured volatility in absolute terms. In fact,
Bachelier used the normalization $H=\frac{\sigma^{B}}{\sqrt{2\pi}}$ and called
this quantity the \textquotedblleft coefficient of
instability\textquotedblright\ or of \textquotedblleft
nervousness\textquotedblright\ of the security $S$. The reason for the
normalisation $H=\frac{\sigma^{B}}{\sqrt{2\pi}}$ is that $H\sqrt{T}$ then
equals the price of an at the money option in Bachelier's model (see \cite{Bac:00})

The Black-Merton-Scholes model (under the risk-neutral measure) for the price
process is, of course, given by
\begin{equation}
S_{t}^{BS}=S_{0}\exp(\sigma^{BS}W_{t}-\frac{(\sigma^{BS})^{2}}{2}
t),\label{blackscholesmodel}
\end{equation}
for $0\leq t\leq T$. Here $\sigma^{BS}$ denotes the usual volatility in the
Black-Merton-Scholes model.

This model was proposed by P. Samuelson in 1965, after he had -- led by an
inquiry of J. Savage for the treatise \cite{Bac:12} -- personally rediscovered
the virtually forgotten Bachelier thesis in the library of Harvard University.
The difference between the two models is somewhat analogous to the difference
between linear and compound interest, as becomes apparent when looking at the
associated It\^{o} stochastic differential equation,
\begin{align*}
dS_{t}^{B} &  =\sigma^{B}dW_{t},\\
dS_{t}^{BS} &  =S_{t}^{BS}\sigma^{BS}dW_{t}.
\end{align*}

This analogy makes us expect that, in the short run, both models should yield
similar results while, in the long run, the difference should be spectacular.
Fortunately, options usually have a relatively short time to maturity (the
options considered by Bachelier had a time to maturity of less than 2 months).

\section{Bachelier versus Black-Merton-Scholes\label{bachelierversusbs}}

We now have assembled all the ingredients to recall the derivation of the
price of an option in Bachelier's framework. Fix a strike price $K$, a horizon
$T$\ and consider the European call $C$, whose pay-off at time $T$ is modeled
by the random variable
\[
C_{T}^{B}=(S_{T}^{B}-K)_{+}.
\]

Applying Bachelier`s \textquotedblleft fundamental principle\textquotedblright
\ and using that $S_{T}^{B}$ is normally distributed with mean $S_{0}$ and
variance $(\sigma^{B})^{2}T$, we obtain for the price of the option at time
$t=0$
\begin{align}
C_{0}^{B} &  =E[(S_{T}^{B}-K)_{+}]\nonumber\\
&  =\int_{K-S_{0}}^{\infty}(S_{0}+x-K)\frac{1}{\sigma^{B}\sqrt{2\pi T}}
\exp(-\frac{x^{2}}{2(\sigma^{B})^{2}T})dx\label{equation}\\
&  =(S_{0}-K)\Phi(\frac{S_{0}-K}{\sigma^{B}\sqrt{T}})+\sigma^{B}\sqrt{T}
\phi(\frac{S_{0}-K}{\sigma^{B}\sqrt{T}}),\label{bachelier}
\end{align}
where $\phi(x)=\frac{1}{\sqrt{2\pi}}\exp(-\frac{x^{2}}{2})$ denotes the
density of the standard normal distribution. We applied the relation
$\phi^{\prime}(x)=-x\phi(x)$ to pass from (\ref{equation}) to (\ref{bachelier}). For details
see, e.g.~, \cite{DelSch:05}.

For further use we shall need the very well-known Black-Merton-Scholes price,
too,
\begin{align}
C_{0}^{BS} &  =E[(S_{T}^{BS}-K)_{+}]\nonumber\\
&  =\int_{-\infty}^{\infty}(S_{0}\exp(-\frac{(\sigma^{BS})^{2}T}{2}
+\sigma^{BS}\sqrt{T}x)-K)_{+}\frac{1}{\sqrt{2\pi}}\exp(-\frac{x^{2}}{2})dx\\
&  =\int_{\frac{\log\frac{K}{S_{0}}+\frac{(\sigma^{BS})^{2}T}{2}}{\sigma
^{BS}\sqrt{T}}}^{\infty}(S_{0}\exp(-\frac{(\sigma^{BS})^{2}T}{2}+\sigma
^{BS}\sqrt{T}x)-K)\frac{1}{\sqrt{2\pi}}\exp(-\frac{x^{2}}{2})dx\\
&  =S_{0}\Phi(\frac{\log\frac{S_{0}}{K}+\frac{1}{2}(\sigma^{BS})^{2}T}
{\sigma^{BS}\sqrt{T}})-K\Phi(\frac{\log\frac{S_{0}}{K}-\frac{1}{2}(\sigma
^{BS})^{2}T}{\sigma^{BS}\sqrt{T}}).\label{bs-price}
\end{align}

Interestingly, Bachelier explicitly wrote down formula (\ref{equation}), but
did not bother to spell out formula (\ref{bachelier}), see \cite[p.
50]{Bac:00}. The main reason seems to be that at his time option prices -- at
least in Paris -- were quoted the other way around: while today the strike
prices $K$ is fixed and the option price fluctuates according to supply and
demand, at Bachelier's times the option prices were fixed (at $10$, $20$ and
$50$ Centimes for a \textquotedblleft rente\textquotedblright, i.e., a
perpetual bond\ with par value of $100$ Francs) and therefore the strike
prices $K$ fluctuated. What Bachelier really needed was the inverse version of
the above relation between the option price $C_{0}^{B}$ and the strike price
$K$.

Apparently there is no simple \textquotedblleft formula\textquotedblright\ to
express this inverse relationship. This is somewhat analogous to the situation
in the Black-Merton-Scholes model, where there is also no \textquotedblleft
formula\textquotedblright\ for the inverse problem of calculating the implied
volatility as a function of the given option price.

We shall see below that L. Bachelier designed a clever series expansion for
$C_{0}^{B}$ as a function of the strike price $K$ in order to derive (very)
easy formulae which \emph{approximate} this inverse relation and which were
well suited to pre-computer technology.

\subsection{At the money options}

Bachelier first specializes to the case of at the money options (called
\textquotedblleft simple options\textquotedblright\ in the terminology of
1900), when $S_{0}=K$. In this case (\ref{bachelier}) reduces to the simple
and beautiful relation
\[
C_{0}^{B}=\sigma^{B}\sqrt{\frac{T}{2\pi}}.
\]
As explicitly noticed by Bachelier, this formula can also be used, for a given
price $C=C_{0}^{B}$\ of an at the money option with maturity $T$, to determine
the \textquotedblleft coefficient of nervousness of the
security\textquotedblright\ $H=\frac{\sigma^{B}}{\sqrt{2\pi}}$, i.e., to
determine the implied volatility in modern language. Indeed, it suffices to
normalize the price $C_{0}^{B}$ by $\sqrt{T}$ to obtain $H=\frac{C_{0}^{B}
}{\sqrt{T}}$. We summarize this fact in the subsequent proposition. For
convenience we phrase it rather in terms of $\sigma^{B}$ than of $H$.

\begin{proposition}
The volatility $\sigma^{B}$\ in the Bachelier model is determined by the price
$C_{0}^{B}$ of an at the money option with maturity $T$ by the relation
\begin{equation}
\sigma^{B}=C_{0}^{B}\sqrt{\frac{2\pi}{T}}.\label{atthemoney}
\end{equation}
\label{proposition at the money}
\end{proposition}

In the subsequent Proposition, we compare the price of an at the money call
option as obtained from the Black-Merton-Scholes and Bachelier's formula
respectively. In order to relate optimally (see the last section)
$C_{0}^{B}$ and $C_{0}^{BS}$ we choose $\sigma^{B}=S_{0}\sigma$ and
$\sigma^{BS}=\sigma$ for some constant $\sigma>0$. We also compare the implied
volatilities, for given price $C_{0}$ of an at the money call with maturity
$T$, in the Bachelier and Black-Merton-Scholes model. We denote the respective
implied volatilities by $\sigma^{B}$ and $\sigma^{BS}$ and discover that the
implied Bachelier volatility estimates the Black-Scholes implied volatility
quite well at the money.

\begin{proposition}
Fix $\sigma>0$, $T>0$ and $S_{0}=K$ (at the money), let $\sigma^{BS}=\sigma$ and $\sigma
^{B}=S_{0}\sigma$ and denote by $C^{B}$ and $C^{BS}$ the corresponding prices
for a European call option in the Bachelier (\ref{bacheliermodel}) and
Black-Merton-Scholes model (\ref{blackscholesmodel}) respectively. Then
\begin{equation}
0\leq C_{0}^{B}-C_{0}^{BS}\leq\frac{S_{0}}{12\sqrt{2\pi}}\sigma^{3}T^{\frac
{3}{2}}=\mathcal{O}((\sigma\sqrt{T})^{3}).\label{price estimation}
\end{equation}
The relative error can be estimated by
\begin{equation}
\frac{C_0^B - C_0^{BS}}{C_0^B} \leq \frac{T}{12} \sigma^2 .
\end{equation}
Conversely, fix the price $0<C_{0}<S_{0}$ of an at the money option and denote
by $\sigma^{B}$ the implied Bachelier volatility and by $\sigma^{BS}$ the
implied Black-Merton-Scholes volatility, then
\begin{equation}
0\leq\sigma^{BS}-\frac{\sigma^{B}}{S_{0}}\leq\frac{T}{12}(\sigma^{BS}
)^{3}.\label{volatility estimation}
\end{equation}
\label{atthemoney estimate}
\end{proposition}

\begin{proof}
(compare \cite{DelSch:05} and \cite{Sch:00}). For $S_{0}=K$, we obtain in the
Bachelier and Black-Merton-Scholes model the following prices, respectively,
\begin{align*}
C_{0}^{B} &  =\frac{S_{0}\sigma}{\sqrt{2\pi}}\sqrt{T}\\
C_{0}^{BS} &  =S_{0}(\Phi(\frac{1}{2}\sigma\sqrt{T})-\Phi(-\frac{1}{2}
\sigma\sqrt{T})).
\end{align*}
Hence
\begin{align*}
C_{0}^{B}-C_{0}^{BS} &  =(\frac{S_{0}}{\sqrt{2\pi}}x-S_{0}(\Phi(\frac{x}
{2})-\Phi(-\frac{x}{2})))|_{x=\sigma\sqrt{T}}\\
&  =\frac{S_{0}}{\sqrt{2\pi}}\left.  \left(  \int_{-\frac{x}{2}}^{\frac{x}{2}
}(1-\exp(-\frac{y^{2}}{2}))dy\right)  \right\vert _{x=\sigma\sqrt{T}}\\
&  \leq\frac{S_{0}}{\sqrt{2\pi}}\int_{-\frac{x}{2}}^{\frac{x}{2}}\frac{y^{2}
}{2}dy|_{x=\sigma\sqrt{T}}\\
&  =\frac{S_{0}}{\sqrt{2\pi}}\frac{x^{3}}{12}|_{x=\sigma\sqrt{T}}=\frac{S_{0}
}{24\sqrt{2\pi}}\sigma^{3}T^{\frac{3}{2}}=\mathcal{O}((\sigma\sqrt{T})^{3}),
\end{align*}
since $e^{y}\geq1+y$ for all $y$, so that $\frac{y^{2}}{2}\geq1-e^{-\frac
{y^{2}}{2}}$ for all $y$. Clearly we obtain $C_{0}^{B}-C_{0}^{BS}\geq0$ again
from the first line.

For the second assertion note that solving equation
\[
C_{0}=\frac{\sigma^{B}}{\sqrt{2\pi}}\sqrt{T}=S_{0}(\Phi(\frac{1}{2}\sigma
^{BS}\sqrt{T})-\Phi(-\frac{1}{2}\sigma^{BS}\sqrt{T}))
\]
for given $\sigma^{B}>0$ yields the Black-Merton-Scholes implied volatility
$\sigma^{BS}$. We obtain similarly as above
\begin{align*}
0 &  \leq\sigma^{BS}-\frac{\sigma^{B}}{S_{0}}=\sigma^{BS}-\frac{\sqrt{2\pi}
}{\sqrt{T}}(\Phi(\frac{1}{2}\sigma^{BS}\sqrt{T})+\Phi(-\frac{1}{2}\sigma
^{BS}\sqrt{T}))\\
&  =\frac{\sqrt{2\pi}}{\sqrt{T}}(\frac{1}{\sqrt{2\pi}}x-(\Phi(\frac{x}
{2})-\Phi(-\frac{x}{2})))|_{x=\sigma^{BS}\sqrt{T}}\\
&  \leq\frac{\sqrt{2\pi}}{\sqrt{T}}\frac{1}{12\sqrt{2\pi}}(\sigma^{BS}
)^{3}T^{\frac{3}{2}}=\frac{(\sigma^{BS})^{3}T}{12}.
\end{align*}

\end{proof}

Proposition \ref{proposition at the money} and \ref{atthemoney estimate} yield
in particular the well-known asymptotic behaviour of an at the money call
price in the Black-Merton-Scholes model for $T\rightarrow\infty$ as described
in \cite{AnaBoy:77}.

Proposition \ref{atthemoney estimate} tells us that for the case when
$(\sigma\sqrt{T})\ll 1$ (which typically holds true in applications), formula
(\ref{atthemoney}) gives a satisfactory approximation of the implied
Black-Merton-Scholes volatility, and is very easy to calculate. We note that
for the data reported by Bachelier (see \cite{Bac:00} and \cite{Sch:00}), the
yearly relative volatility was of the order of $ 2.4 \% $ p.a. and $T$ in the
order of one month, i.e $T=\frac{1}{12}$ years so that $\sqrt{T}\approx0.3$.
Consequently we get $(\sigma\sqrt{T})^{3}\approx(0.008)^{3}\approx
5\times10^{-7}$. The estimate in Proposition \ref{atthemoney estimate} yields
a right hand side of $\frac{S_{0}}{12\sqrt{2\pi}}5\times10^{-7} \approx 1.6\times
10^{-8}S_{0}$, i.e. the difference of the Bachelier and Black-Merton-Scholes
price (when using the same volatility $\sigma=2.4 \% $ p.a.) is of the order
$10^{-8}$ of the price $S_{0}$ of the underlying security.

The above discussion pertains to the limiting behaviour, for $ \sigma \sqrt{T} \to 0 $, of 
the Bachelier and Black-Merton-Scholes prices of at the money options, i.e.~, where $S_0=K$. One might ask the same question for the case $ S_0 \neq K $. Unfortunately, this question turns out not to be meaningful from a financial point of view. Indeed, if we fix $ S_0 \neq K $ and let 
$ \sigma \sqrt{T} $ tend to zero, then the Bachelier price as well as the Black-Scholes price
tend to the pay-off function $ {(S_0 - K)}_+ $ of order higher than $ {(\sigma \sqrt{T})}^n $, for every $ n \geq 1 $. Hence, in particular, their difference tends to zero faster than any power of $ (\sigma \sqrt{T}) $. This does not seem to us an interesting result and corresponds -- in financial terms -- to the fact that, for fixed $ S_0 \neq K $ and letting
$ \sigma \sqrt{T} $ tend to zero, the hinked pay-off function $ {(S_0 - K)}_+ $ essentially amounts to the same as the linear pay-off functions $ S_0 - K $, in the case $ S_0 > K $, and $ 0 $, in the case $ S_0 - K < 0 $. In particular the functional dependence of the prices on $ \sigma \sqrt{T} $ is not analytical. This behaviour is essentially different from the behaviour at the money, $ S_0 = K $, since there the prices depend analytically on $ \sigma \sqrt{T} $ and the difference tends to zero of order $ 3 $.

\section{Further results of L. Bachelier}

We now proceed to a more detailed analysis of the option pricing formula
(\ref{bachelier}) for general strike prices $K$. Let $C=C_{0}^{B}$ denote the
option price from (\ref{bachelier}). We shall introduce some notation used by
L. Bachelier for the following two reasons: firstly, it should make the task
easier for the interested reader to look up the original texts by Bachelier;
secondly, and more importantly, we shall see that his notation has its own
merits and allows for intuitive and economically meaningful interpretations
(as we have already seen for the normalization $H=\frac{\sigma^{B}}{\sqrt
{2\pi}}$ of the volatility, which equals the time-standardized price of an at
the money option).

L. Bachelier found it convenient to use a parallel shift of the coordinate
system moving $S_{0}$ to $0$,\ so that the Gaussian distribution will be
centered at $0$. We write
\begin{equation}
a=\frac{\sigma^{B}\sqrt{T}}{\sqrt{2\pi}},\quad m:=K-S_{0},\quad P:=m+C.
\label{notation}
\end{equation}

The parameter $a$ equals, up to the normalizing factor $\frac{S_{0}}
{\sqrt{2\pi}}$, the \emph{time standardized absolute volatility }$\sigma
^{B}\sqrt{T}$ \emph{at maturity }$T$. Readers familiar, e.g. with the
Hull-White model of stochastic volatility, will realize that this is a very
natural parametrization for an option with maturity $T$.

In any case, the quantity $a$ was a natural parametrization for L. Bachelier,
as it is the \emph{price of the at the money option} with the maturity $T$
(see formula \ref{atthemoney}), so that it can be directly observed from market data.

The quantity $m$ is the difference between the strike price $K$ and $S_{0}$ and needs no
further explanation. $P$ has a natural interpretation (in Bachelier's times it
was called \textquotedblleft\'{e}cart\textquotedblright, i.e. the
\textquotedblleft spread\textquotedblright\ of an option): it is the price $P$
of a european put with maturity $T$ and strike price $K$, as was explicitly
noted by Bachelier (using, of course, different terminology). In today's terminology this amounts to the put-call parity. Bachelier interpreted $P$ as the premium of an insurance against prices falling below
$K$.

This is nicely explained in \cite{Bac:00}: a speculator \textquotedblleft\'{a}
la hausse\textquotedblright, i.e. hoping for a rise of $S_{T}$ may buy a
forward contract with maturity $T$. Using the \textquotedblleft fundamental
principle\textquotedblright, which in this case boils down to elementary no
arbitrage arguments, one concludes that the forward price must equal $S_{0}$,
so that the total gain or loss of this operation is given by the random
variable $S_{T}-S_{0}$ at time $T$.

On the other hand , a more prudent speculator might want to limit the maximal
loss by a quantity $K>0$.\ She thus would buy a call option with price $C$,
which would correspond to a strike price $K$ (here we see very nicely the
above mentioned fact that in Bachelier`s times the strike price was considered
as a function of the option price $C$ -- la \textquotedblleft
prime\textquotedblright\ in french -- and not vice versa as today). Her total
gain or loss would then be given by the random variable
\[
(S_{T}-K)_{+}-C.
\]

If at time $T$ it indeed turns out that $S_{T}\geq K$, then the buyer of the
forward contract is, of course, better off than the option buyer. The
difference equals
\[
(S_{T}-S_{0})-[(S_{T}-K)-C]=K-S_{0}+C=P,
\]
which therefore may be interpreted as a \textquotedblleft cost of
insurance\textquotedblright. If $S_{T}\leq K$, we obtain
\[
(S_{T}-S_{0})-[0-C]=(S_{T}-K)+K-S_{0}+C=(S_{T}-K)+P.
\]
By the Bachelier's \textquotedblleft fundamental principle\textquotedblright
\ we obtain
\[
P=E[(S_{T}-K)_{-}].
\]
Hence Bachelier was led by no-arbitrage considerations to the put-call parity.
For further considerations we denote the put price in the Bachelier model at
time $t=0$ by $P_{0}^{B}$ or $P(m)$ respectively. Clearly, the higher the
potential loss $C$ is, which the option buyer is ready to accept, the lower
the costs of insurance $P$\ should be and vice versa, so that we expect a
monotone dependence of these two quantities.

In fact, Bachelier observed that the following pretty result holds true in his
model (see \cite{Bac:12}, p.295):

\begin{proposition}
[Theorem of reciprocity]For fixed $\sigma^{B}>0$ and $T>0$ the quantities $C$
and $P$ are reciprocal in Bachelier's model, i.e. there is a monotone,
strictly decreasing and self-inverse (that is $I=I^{-1}$) function
$I:\mathbb{R}_{>0}\rightarrow\mathbb{R}_{>0}$ such that $P=I(C)$
.\label{reciprocity}
\end{proposition}

\begin{proof}
Denote by $\psi$ the density of $S_{T}-S_{0}$, then
\begin{align*}
C(m)  &  =\int_{m}^{\infty}(x-m)\psi(x)dx,\\
P(m)  &  =\int_{-\infty}^{m}(m-x)\psi(x)dx.
\end{align*}
Hence we obtain that $C(-m)=P(m)$. We note in passing that this is only due to
the symmetry of the density $\psi$ with respect to reflection at $0$. Since
$C^{\prime}(m)<0$ (see the proof of Proposition \ref{expansion}) we obtain
$P=P(m(C)):=I(C)$, where $C\mapsto m(C)$ inverts the function $m\mapsto C(m)$.
$C$ maps $\mathbb{R}$ in a strictly decreasing way to $\mathbb{R}_{>0}$ and
$P$ maps $\mathbb{R}$ in a strictly increasing way to $\mathbb{R}_{>0}$. The
resulting map $I$ is therefore strictly decreasing, and -- due to symmetry --
we obtain
\[
I(P)=P(m(P))=P(-m(C))=C,
\]
so $I$ is self-inverse.
\end{proof}

Using the above notations (\ref{notation}), equation (\ref{bachelier}) for the
option price $C_{0}^{B}$ (which we now write as $C(m)$ to stress the
dependence on the strike price) obtained from the fundamental principle
becomes
\begin{equation}
C(m)=\int_{m}^{\infty}(x-m)\mu(dx), \label{integralequation}
\end{equation}
where $\mu$ denotes the distribution of $S_{T}-S_{0}$, which has the Gaussian
density $\mu(dx)=\psi(x)dx$,
\[
\psi(x)=\frac{1}{\sigma^{B}\sqrt{2\pi T}}\exp(-\frac{x^{2}}{2(\sigma^{B}
)^{2}T})=\frac{1}{2\pi a}\exp(-\frac{x^{2}}{4\pi a^{2}}).
\]

As mentioned above, Bachelier does not simply calculate the integral
(\ref{integralequation}). He rather does something more interesting (see
\cite[p. 294]{Bac:12}): \textquotedblleft Si l'on d\'{e}veloppe l'integrale en
s\'{e}rie, on obtient \textquotedblright, i.e. "if one develops the integral
into a series one obtains..."
\begin{equation}
C(m)=a-\frac{m}{2}+\frac{m^{2}}{4\pi a}-\frac{m^{4}}{96\pi^{2}a^{3}}
+\frac{m^{6}}{1920\pi^{3}a^{5}}+\dots^{\prime\prime}.\label{original formula bachelier}
\end{equation}
In the subsequent theorem we justify this step. It is worth noting that the method
for developing this series expansion is not restricted to Bachelier's model,
but holds true in general (provided that $\mu$, the probability distribution
of $S_{T}-S_{0}$, admits a density function $\psi$, which is analytic in a
neighborhood of $0$).

\begin{theorem}
Suppose that the law $\mu$ of the random variable $S_{T}$ admits a density
\[
\mu(dx)=\psi(x)dx,
\]
such that $\psi$ is analytic in a ball of radius $r>0$ around $0$, and that
\[
\int_{-\infty}^{\infty}x\psi(x)dx<\infty.
\]
Then the function
\[
C(m)=\int_{m}^{\infty}(x-m)\mu(dx)
\]
is analytic for $|m|<r$ and admits a power series expansion
\[
C(m)=\sum_{k=0}^{\infty}c_{k}m^{k},
\]
where $c_{0}=\int_{0}^{\infty}x\psi(x)dx$, $c_{1}=-\int_{0}^{\infty}\psi(x)dx$
and $c_{k}=\frac{1}{k!}\psi^{(k-2)}(0)$ for $k\geq2$.\label{expansion}
\end{theorem}

\begin{proof}
Due to our assumptions $C$ is seen to be analytic as sum of two analytic
functions,
\[
C(m)=\int_{m}^{\infty}x\psi(x)dx-m\int_{m}^{\infty}\psi(x)dx.
\]
Indeed, if $\psi$ is analytic around $0$, then the functions $x\mapsto
x\psi(x)$ and $m\mapsto\int_{m}^{\infty}x\psi(x)dx$ are analytic with the same
radius of convergence $r$. The same holds true for the function $m\mapsto
m\int_{m}^{\infty}\psi(x)dx$. The derivatives can be calculated by the Leibniz
rule,
\begin{align*}
C^{\prime}(m)  &  =-m\psi(m)-\int_{m}^{\infty}\psi(x)dx+m\psi(m)\\
&  =-\int_{m}^{\infty}\psi(x)dx,\\
C^{\prime\prime}(m)  &  =\psi(m),
\end{align*}
whence we obtain for the $k$-th derivative,
\[
C^{(k)}(m)=\psi^{(k-2)}(m),
\]
for $k\geq2$.
\end{proof}

\begin{remark}
If we assume that $m\mapsto C(m)$ is locally analytic around $m=0$ (without
any assumption on the density $\psi$), then the density $x\mapsto\psi(x)$ is
analytic around $x=0$, too, by inversion of the above argument.
\end{remark}

\begin{remark}
In the case when $\psi$ equals the Gaussian distribution, the calculation of
the Taylor coefficients yields
\begin{align*}
\frac{d}{dy}(\frac{1}{2\pi a}\exp(-\frac{y^{2}}{4\pi a^{2}}))  &  =-\frac
{1}{4}\frac{y}{\pi^{2}a^{3}}e^{-\frac{1}{4}\frac{y^{2}}{\pi a^{2}}},\\
\frac{d^{2}}{dy^{2}}(\frac{1}{2\pi a}\exp(-\frac{y^{2}}{4\pi a^{2}}))  &
=-\frac{1}{8}\frac{2\pi a^{2}-y^{2}}{\pi^{3}a^{5}}e^{-\frac{1}{4}\frac{y^{2}
}{\pi a^{2}}},\\
\frac{d^{3}}{dy^{3}}(\frac{1}{2\pi a}\exp(-\frac{y^{2}}{4\pi a^{2}}))  &
=\frac{1}{16}\frac{6\pi a^{2}y-y^{3}}{\pi^{4}a^{7}}e^{-\frac{1}{4}\frac{y^{2}
}{\pi a^{2}}},\\
\frac{d^{4}}{dy^{4}}(\frac{1}{2\pi a}\exp(-\frac{y^{2}}{4\pi a^{2}}))  &
=\frac{1}{32}\frac{12\pi^{2}a^{4}-12y^{2}\pi a^{2}+y^{4}}{\pi^{5}a^{9}
}e^{-\frac{1}{4}\frac{y^{2}}{\pi a^{2}}},
\end{align*}
Consequently $\psi(0)=\frac{1}{2\pi a}$, $\psi^{\prime}(0)=0$, $\psi
^{\prime\prime}(0)=-\frac{1}{4\pi^{2}a^{3}}$, $\psi^{\prime\prime\prime}(0)=0$
and $\psi^{\prime\prime\prime\prime}(0)=\frac{3}{8}\frac{1}{\pi^{3}a^{5}}$,
hence with $C(0)=a$ and $C^{\prime}(0)=-\frac{1}{2}$,
\begin{equation}
C(m)=a-\frac{m}{2}+\frac{m^{2}}{4\pi a}-\frac{m^{4}}{96\pi^{2}a^{3}}
+\frac{m^{6}}{1920\pi^{3}a^{5}}+\mathcal{O}(m^{8}) \label{call-expansion}
\end{equation}
and the series converges for all $m$, as the Gaussian distribution is an
entire function. This is the expansion indicated by Bachelier in
\cite{Bac:12}. Since $P(-m)=C(m)$, we also obtain the expansion for the put
\begin{equation}
P(m)=a+\frac{m}{2}+\frac{m^{2}}{4\pi a}-\frac{m^{4}}{96\pi^{2}a^{3}}
+\frac{m^{6}}{1920\pi^{3}a^{5}}+\mathcal{O}(m^{8}). \label{put-expansion}
\end{equation}

\end{remark}

\begin{remark}
Looking once more at Bachelier's series one notes that it is rather a Taylor
expansion in $\frac{m}{a}$ than in $m$. Note furthermore that $\frac{m}{a}$ is
a dimensionless quantity. The series then becomes
\begin{align*}
C(m)  &  =a~F(\frac{m}{a})\\
F(x)  &  =1-\frac{x}{2}+\frac{x^{2}}{4\pi}-\frac{x^{4}}{96\pi^{2}}+\frac
{x^{6}}{1920\pi^{3}}+\mathcal{O}(x^{8}).
\end{align*}
We note as a curiosity that already in the second order term the number $\pi$
appears. Whence -- if we believe in Bachelier's formula for option pricing --
we are able to determine $\pi$ at least approximately (see
(\ref{quadraticapprox}) below) -- from financial market data (compare Georges
Louis Leclerc Comte de Buffon's method to determine $\pi$ by using statistical experiments).
\end{remark}

Let us turn back to Bachelier's original calculations. He first truncated the
Taylor series (\ref{call-expansion}) after the quadratic term, i.e.
\begin{equation}
C(m)\approx a-\frac{m}{2}+\frac{m^{2}}{4\pi a}. \label{quadraticapprox}
\end{equation}
This (approximate) formula can easily be inverted by solving a quadratic
equation, thus yielding an explicit formula for $m$ as a function of $C$.
Bachelier observes that the approximation works well for small values of
$\frac{m}{a}$\ (the cases relevant for his practical applications) and gives
some numerical estimates. We summarize the situation.

\begin{proposition}
[Rule of Thumb 1]For given maturity $T$, strike $K$ and $\sigma^{B}>0$,
let $m=K-S_{0}$ and denote by $a=C(0)$ the Bachelier price of the at the
money option and by $C(m)$ the Bachelier price of the call option with strike
$K=S_{0}+m$. Define
\begin{align}
\widehat{C}(m) &  =a-\frac{m}{2}+\frac{m^{2}}{4\pi a}\label{bacheliersapprox}
\\
&  =C(0)-\frac{m}{2}+\frac{m^{2}}{4\pi C(0)},
\end{align}
then we get an approximation of the Bachelier price $C(m)$ of order $4$, i.e.
$C(m)-\widehat{C}(m)=\mathcal{O}(m^{4})$.
\end{proposition}

\begin{remark}
Note that the value of $\widehat{C}(m)$ only depends on the price $a=C(0)$ of
an at the money option (which is observable at the market) and the given
quantity $m=K-S_{0}$.
\end{remark}

\begin{remark}
Given any stock price model under a risk neutral measure, the above approach
of quadratic approximation can be applied if the density $\psi$ of
$S_{T}-S_{0}$ is locally analytic and admits first moments. The approximation
then reads
\begin{equation}
\widehat{C}(m)=C(0)-B(0)m+\frac{\psi(0)}{2}m^{2}\label{rule of thumb 1}
\end{equation}
up to a term of order $\mathcal{O}(m^{3})$. Here $C(0)$ denotes the price of
the at the money european call option (pay-off $(S_{T}-S_{0})_{+}$), $B(0)$
the price of the at the money binary option (pay-off $1_{\{S_{T}\geq S_{0}\}}
$) and $\psi(0)$ the value of an at the money \textquotedblleft
Dirac\textquotedblright\ option (pay-off $\delta_{S_{0}}$, with an appropriate
interpretation as a limit). Notice that $ B(0) $ can also be interpreted, in Bachelier's model, as
the hedging ration Delta.
\end{remark}

\begin{example}
Take for instance the Black-Merton-Scholes model, then the terms of the
quadratic approximation (\ref{rule of thumb 1}) can be calculated easily,
\begin{align*}
C(0) &  =S_{0}(\Phi(\frac{1}{2}\sigma^{BS}\sqrt{T})-\Phi(-\frac{1}{2}
\sigma^{BS}\sqrt{T}))\\
B(0) &  =\Phi(-\frac{1}{2}\sigma^{BS}\sqrt{T}),\\
\psi(0) &  =\frac{1}{\sigma^{BS}\sqrt{2\pi T}}\frac{1}{S_{0}}\exp(-\frac{1}
{8}(\sigma^{BS})T).
\end{align*}
Notice that the density $\psi$ of $S_{T}-S_{0}$ in the Black-Merton-Scholes
model is not an entire function, hence the Taylor expansion only converges
with a finite radius of convergence.
\end{example}

Although Bachelier had achieved with formula (\ref{bacheliersapprox}) a
practically satisfactory solution, which allowed to calculate (approximately)
$m$ as a function of $C$ by only using pre-computer technology, he was not
entirely satisfied. Following the reflexes of a true mathematician he tried to
obtain better approximations (yielding still easily computable quantities)
than simply truncating the Taylor series after the quadratic term. He observed
that, using the series expansion for the put option (see formula
\ref{put-expansion})
\[
P(m)=a+\frac{m}{2}+\frac{m^{2}}{4\pi a}-\frac{m^{4}}{96\pi^{2}a^{3}}+...
\]
and computing the product function $C(m)P(m)$\ or, somewhat more
sophisticatedly, the triple product function $C(m)P(m)\frac{C(m)+P(m)}{2}$,
one obtains interesting cancellations in the corresponding Taylor series,
\begin{align*}
C(m)P(m)  &  =a^{2}-\frac{m^{2}}{4}+\frac{m^{2}}{2\pi}+\mathcal{O}(m^{4}),\\
C(m)P(m)\frac{C(m)+P(m)}{2}  &  =a^{3}-\frac{m^{2}a}{4}+\frac{3m^{2}a}{4\pi
}+\mathcal{O}(m^{4}).
\end{align*}
Observe that $(C(m)P(m))^{\frac{1}{2}}$ is the geometric mean of the
corresponding call and put price, while $(C(m)P(m)\frac{C(m)+P(m)}{2}
)^{\frac{1}{3}}$ is the geometric mean of the call, the put and the arithmetic
mean of the call and put price.

The latter equation yields the approximate identity
\[
(C(m)+P(m))C(m)P(m)\approx2a^{3}
\]
which Bachelier rephrases as a cooking book recipe (see \cite{Bac:12}, p.201):
\begin{align*}
&  \text{On additionne l'importance de la prime et son \'{e}cart.}\\
&  \text{On multiplie l'importance de la prime par son \'{e}cart.}\\
&  \text{On fait le produit des deux r\'{e}sultats.}\\
&  \text{Ce produit doit \^{e}tre le m\^{e}me pour toutes les primes}\\
&  \text{qui ont m\^{e}me \'{e}ch\'{e}ance,}
\end{align*}
i.e. "One adds up the call and put price, one multiplies the call and the put
price, one multiplies the two results. The product has to be the same
for all premia, which correspond to options of the same maturity $a$." This recipe allows to
approximately calculate for $m\neq m^{\prime}$\ in a quadruple
\[
(C(m),P(m),C(m^{\prime}),P(m^{\prime}))
\]
any one of these four quantities as an easy (from the point of view of
pre-computer technology) function of the other three. Note that, in the case
$m=0$, we have $C(0)=P(0)=a$, which makes the resulting calculation even easier.

We now interpret these equations in a more contemporary language (but, of
course, only rephrasing Bachelier`s insight in this way).

\begin{proposition}
[Rule of Thumb 2]For given $T>0$, $\sigma^{B}>0$ and $m=K-S_{0}$ denote by
$C(m)$\ and $P(m)$ the prices of the corresponding call and put options in the
Bachelier model. Denote by
\begin{align*}
a(m)  &  :=C(m)P(m)\\
b(m)  &  =C(m)P(m)(\frac{C(m)+P(m)}{2})
\end{align*}
the products considered by Bachelier, then we have $a(0)=a^{2}$ and
$b(0)=a^{3}$, and
\begin{align*}
\frac{a(m)}{a(0)}  &  =1-\frac{(\pi-2)(\frac{m}{a})^{2}}{4\pi}+\mathcal{O}
(m^{4}),\\
\frac{b(m)}{b(0)}  &  =1-\frac{(\pi-3)(\frac{m}{a})^{2}}{4\pi}+\mathcal{O}
(m^{4}).
\end{align*}

\end{proposition}

\begin{remark}
We rediscover an approximation of the reciprocity relation of Proposition
\ref{notation} in the first of the two rules of the thumb. Notice also that
this rule of thumb only holds up to order $(\frac{m}{a})^{2}$. Finally note
that $\pi-3\approx0.1416$ while $\pi-2\approx1.1416$, so that the coefficient
of the quadratic term in the above expressions is smaller for $\frac
{b(m)}{b(0)}$ by a factor of $8$ as compared to $\frac{a(m)}{a(0)}$. This is
why Bachelier recommended this slightly more sophisticated product.
\end{remark}

\section{Bachelier versus other models}

Bachelier's model yields very good approximation results with respect to the Black-Scholes-model for at the money options.
This corresponds to notions of weak (absolute or relative) errors of approximation in short time asymptotics, i.e.~ estimates of the quantitiy
$$
|E(f(S_T^B)) - E(f(S_T^{BS}))|
$$
for $ T \to 0 $. Compare \cite{KloPla:92} for the notion of weak and strond errors of approximation in the realm of numerical methods for SDE. This can be very delicate as seen in Proposition \ref{atthemoney estimate} and the remarks thereafter. However, in this section we concentrate on the easier notion of $L^2$-strong errors, which means estimates on the $ L^2 $-distance between models at given points in time $ T $.

We ask two questions: first, how to generalize Bachelier's model in order to obtain better approximations of
the Black-Scholes-model and second, how to extend this approach beyond the Black-Scholes model. Both questions can be answered by the methods from chaos expansion.
There are possible extensions of the Bachelier model in several directions,
but extensions, which improve -- in an optimal way -- the $L^{2}$-distance and
the short-time asymptotics (with respect to a given model), are favorable from the point of view
of applications. This
observation in mind we aim for best (in the sense of $L^{2}$-distance)
approximations of a given general process $(S_{t})_{0\leq t\leq T}$ by
iterated Wiener-Ito integrals up to a certain order. We demand that the
approximating processes are martingales to maintain no arbitrage properties.
The methodology results into the one of chaos expansion or Stroock-Taylor
Theorems (see \cite{Mal:97}). Methods from chaos expansion for the (explicit)
construction of price processes have already proved to be very useful, see for \cite{MalTha:05}.
In the sequel we shall work on a probability space $ (\Omega,\mathcal{F},P) $ carrying a one-dimensional Brownian motion $ {(W_t)}_{0 \leq t \leq T} $ with its natural filtration $ {(\mathcal{F}_t)}_{0 \leq t \leq T} $. For all necessary details on Gaussian spaces, chaos expansion, $n$-th Wiener chaos, etc, see \cite{MalTha:05}. Certainly, the following considerations easily generalize to multi-dimensional Brownian motions.

We shall call any Gaussian martingale in this setting a \emph{(general) Bachelier model}.

\begin{definition}
Fix $N\geq1$. Denote by $(M_{t}^{(n)})_{0\leq t\leq T}$ martingales
with continuous trajectories for $0\leq n\leq N$, such that $M_{t}^{(n)}
\in\mathcal{H}_{n}$ for $0\leq t\leq T$, where $\mathcal{H}_{n}$
denotes the $n$-th Wiener chaos in $L^{2}(\Omega)$. Then we call the process
\[
S_{t}^{(N)}:=\sum_{n=0}^{N}M_{t}^{(n)}
\]
an\emph{ extension of degree }$N$\emph{ of the Bachelier model}. Note that
$M_{t}^{(0)}=M^{(0)}$ is constant, and that $S_{t}^{(1)}=M^{(0)}+M_{t}^{(1)}$
is a (general) Bachelier model.
\end{definition}

Given an $L^{2}$-martingale $(S_{t})_{0\leq t\leq T}$ with $S_{0}>0$
and $N\geq1$. There exists a unique extension of degree $N$ of the Bachelier
model (in the sense that the martingales $(M_{t}^{(n)})_{0\leq t\leq T
}$ are uniquely defined for $0\leq n\leq N$) minimizing the $L^{2}$-norm
$E[(S_{t}-S_{t}^{(N)})^{2}]$ for all $0\leq t\leq T$. Furthermore
$S_{t}^{(N)}\rightarrow S_{t}$ in the $L^{2}$-norm as $N\rightarrow\infty$,
uniformly for $0\leq t\leq T$.

Indeed, since the orthogonal projections $p_{n}:L^{2}(\Omega,\mathcal{F}
_{T},P)\rightarrow L^{2}(\Omega,\mathcal{F}_{T},P)$ onto the
$n$-th Wiener chaos commute with conditional expectations $E(.|\mathcal{F}
_{t})$ (see for instance \cite{Mal:97}), we obtain that
\[
M_{t}^{(n)}:=p_{n}(S_{t}),
\]
for $0\leq t\leq T$, is a martingale with continuous trajectories,
because
\[
E(p_{n}(S_{T})|\mathcal{F}_{t})=p_{n}(E(S_{T}|\mathcal{F}
_{t})=p_{n}(S_{t})
\]
for $0\leq t\leq T$. Consequently
\[
S_{t}^{(N)}:=\sum_{n=0}^{N}M_{t}^{(n)}
\]
is a process minimizing the distance to $S_{t}$ for any $t\in [0,T]$. Clearly we have that
\[
S_{t}:=\sum_{n=0}^{\infty}M_{t}^{(n)}
\]
in the $L^{2}$-topology.

\begin{example}
For the Black-Merton-Scholes model with $\sigma^{BS}=\sigma$ we obtain that
\[
M_{t}^{(n)}=S_{0}\sigma^{n}t^{\frac{n}{2}}H_{n}(\frac{W_{t}}{\sqrt{t}}),
\]
where $H_{n}$ denotes the $n$-th Hermite polynomial, i.e. $(n+1)H_{n+1}
(x)=xH_{n}(x)-H_{n-1}(x)$ and $H_{0}(x)=1$, $H_{1}(x)=x$ for $n\geq1$ (for
details see \cite{Nua:95}). Hence
\begin{align*}
M_{t}^{(0)} &  =S_{0},\\
M_{t}^{(1)} &  =S_{0}\sigma W_{t},\\
M_{t}^{(2)} &  =S_{0}\frac{\sigma^{2}}{2}(W_{t}^{2}-t).
\end{align*}
We recover Bachelier's model as extension of degree $1$ minimizing the
distance to the Black-Merton-Scholes model. Note that we have
\[
||S_{t}-S_{t}^{(N)}||_{2}\leq C_N S_{0}\sigma^{N+1}t^{\frac{N+1}{2}},
\]
for $0\leq t\leq T$. Furthermore we can calculate a sharp constant $C_N$,
namely
\begin{align*}
C_N^{2} &  =\sup_{0\leq t\leq T}\sum_{m\geq N+1}{(\sigma \sqrt{t})}^{2(m-N-1)}E[(H_{m}
(\frac{W_{t}}{\sqrt{t}}))^{2}]\\
&  =\sum_{m\geq N+1} {(\sigma \sqrt{t})}^{2(m-N-1)}\frac{1}{m!}.
\end{align*}
since $E[(H_{m}(\frac{W_{t}}{\sqrt{t}}))^{2}]=\frac{1}{m!}$ for $0<t\leq
T$ and $m\geq0$.\label{blackscholesmodelextension}
\end{example}

Due to the particular structure of the chaos decomposition we can prove the
desired short-time asymptotics:

\begin{theorem}
Given an $L^{2}$-martingale $(S_{t})_{0\leq t\leq T}$, assume that
$S_{T}=\sum_{i=0}^{\infty}W_{i}(f_{i})$ with symmetric $L^{2}
$-functions $f_{i}:[0,T]^{i}\rightarrow\mathbb{R}$ and iterated
Wiener-Ito integrals
\[
W_{t}^{i}(f_{i}):=\int_{0\leq t_{1}\leq\dots\leq t_{i}\leq t}f_{i}(t_{1}
,\dots,t_{i})dW_{t_{1}}\cdots dW_{t_{i}}.
\]
If there is an index $i_{0}$ such that for $i\geq i_{0}$ the functions $f_{i}$
are bounded and $K^{2}:=\sum_{i\geq i_{0}}\frac{(T)^{i-i_{0}}}
{i!}||f_{i}||_{\infty}^{2}<\infty$, then we obtain $||S_{t}-S_{t}^{(N)}
||_{2}\leq Kt^{\frac{n+1}{2}}$ for each $N\geq i_{0}$ and $0\leq t\leq
T$.
\end{theorem}

\begin{proof}
We apply that $E[(E(W_{T}^{i}(f_{i})|\mathcal{F}_{t}))^{2}]=\frac
{1}{i!}||1_{[0,t]}^{\otimes i}f_{i}||_{L^{2}([0,T]^{i})}^{2}\leq
\frac{t^{i}}{i!}||f_{i}||_{\infty}^{2}$ for $i\geq i_{0}$. Hence the result by
applying
\[
||S_{T}-S_{T}^{(N)}||_{2}\leq\sum_{i\geq i_{0}}E[(E(W_{T}^{i}
(f_{i})|\mathcal{F}_{T}))^{2}]\leq KT^{\frac{N+1}{2}}.
\]

\end{proof}

\begin{remark}
Notice that the Stroock-Taylor Theorem (see \cite{Mal:97}, p.161) tells that
for $S_{T}\in\mathcal{D}^{2,\infty}$ the series
\[
\sum_{i=0}^{\infty}W_{T}^{i}((t_{1},\dots,t_{i})\mapsto E(D_{t_{1}
,\dots,t_{i}}S_{T}))=S_{T}
\]
converges in $\mathcal{D}^{2,\infty}$. Hence the above condition is a
statement about boundedness of higher Malliavin derivatives. The well-known
case of the Black-Merton-Scholes model yields
\[
D_{t_{1},\dots,t_{i}}S_{T}^{BS}=1_{[0,T]}^{\otimes i}
\]
for $i\geq0$, so the condition of the previous theorem is satisfied.
\end{remark}

\section{Appendix}

We provide tables with Bachelier and Black-Scholes implied volatilities for different times to maturity, where we have used Nasdaq-index-option data in discounted prices in order to compare the results. Circles represent the Black-Scholes implied
volatility of data points above a certain level of trade volume. The dotted line represents the implied Bachelier (relative) volatility
$ \sigma^B = S_0 \sigma_{\emph{rel}} $.

\begin{center}
	\includegraphics[angle=270,width=1.00\textwidth]{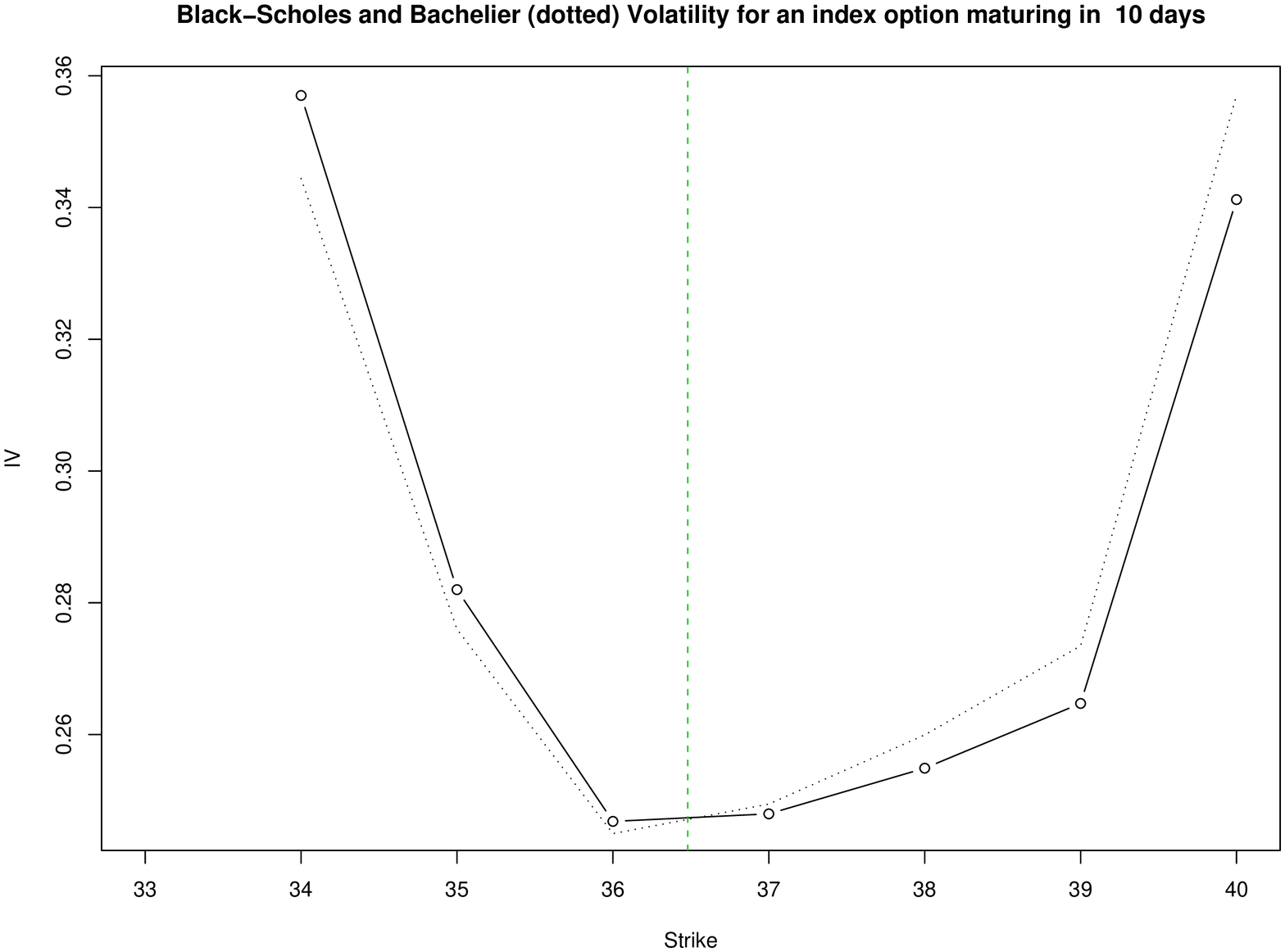}
\end{center}

\begin{center}
	\includegraphics[angle=270,width=1.00\textwidth]{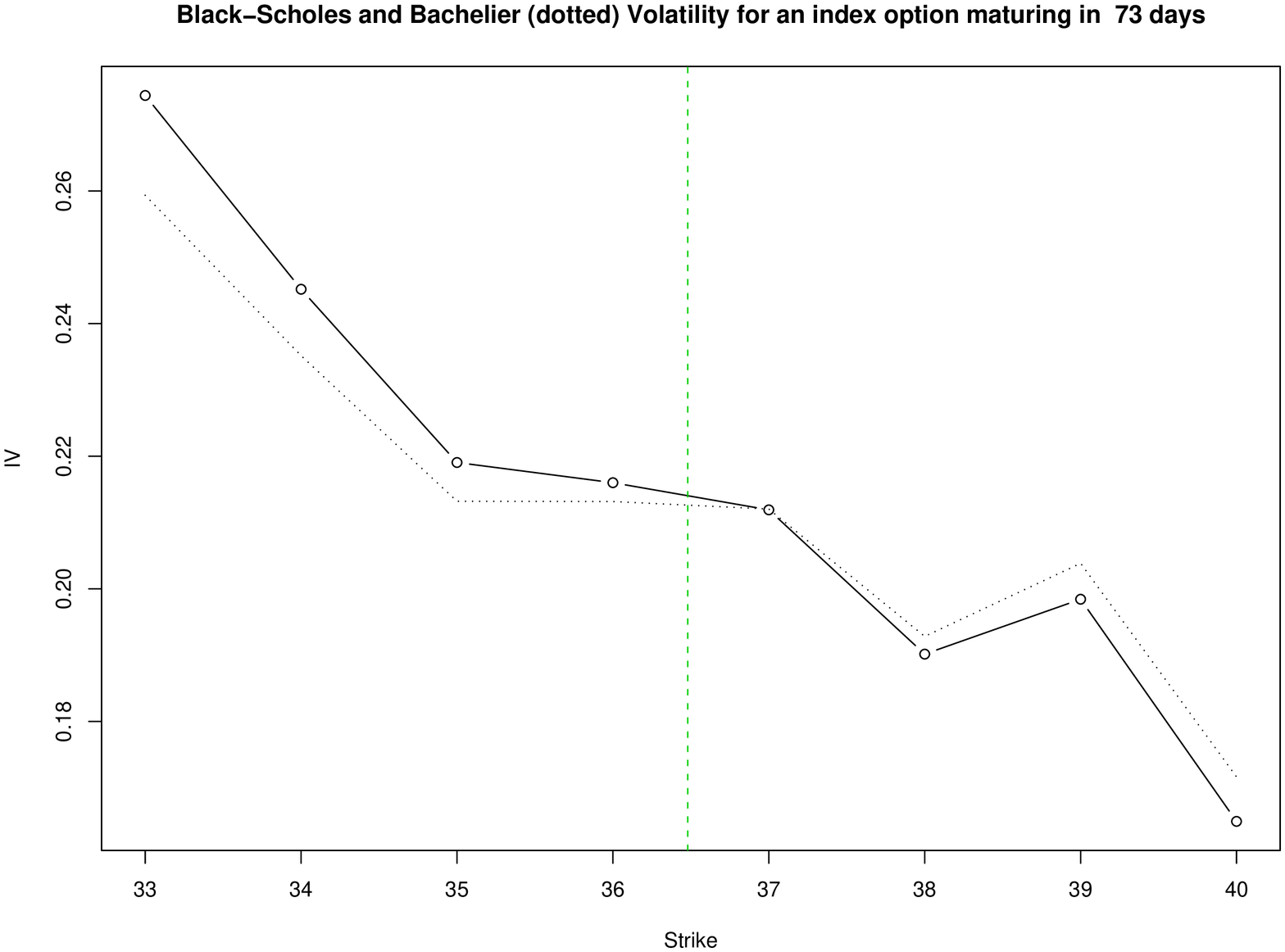}
\end{center}

\begin{center}
	\includegraphics[angle=270,width=1.00\textwidth]{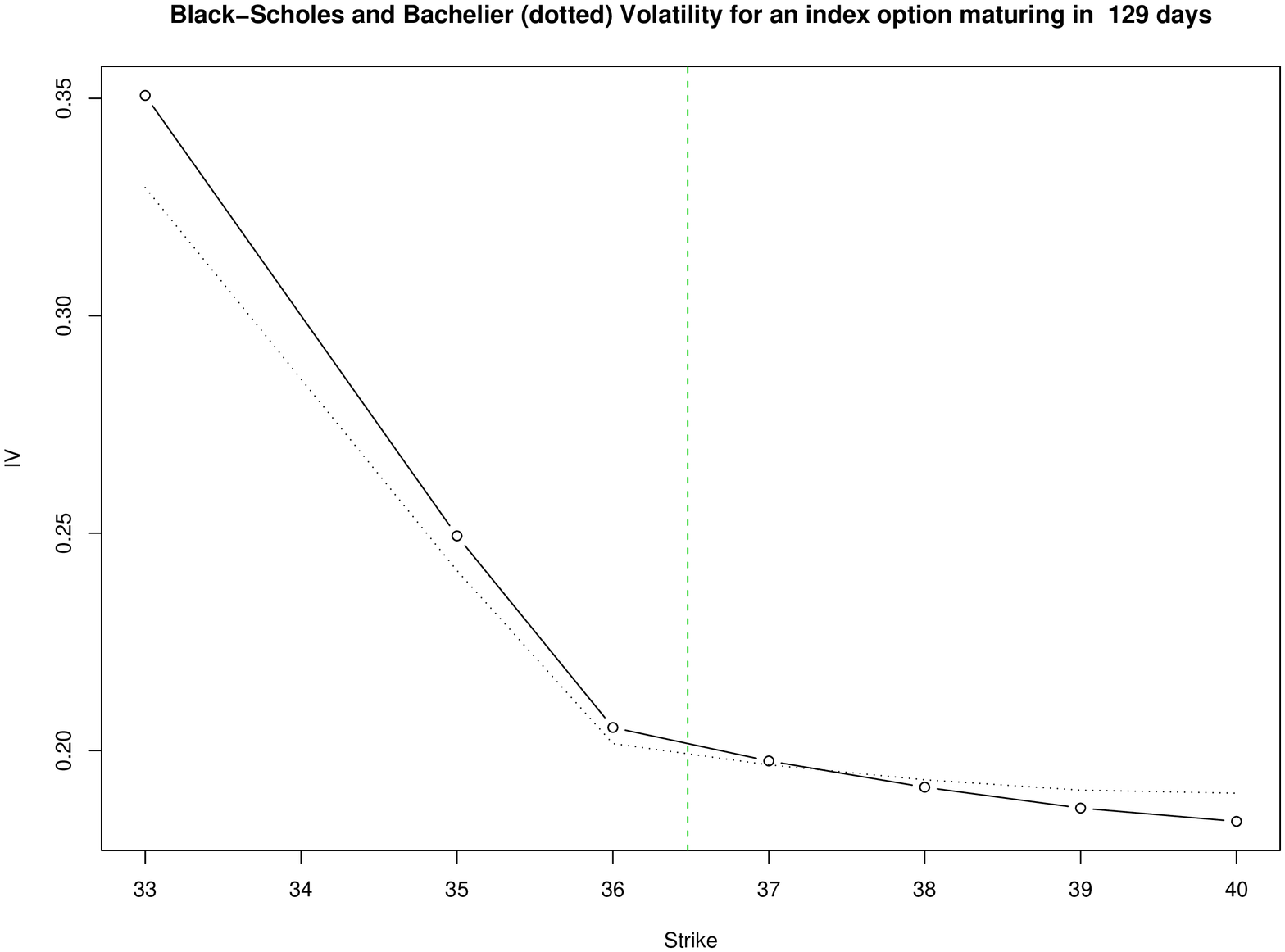}
\end{center}
\end{document}